\documentclass[preprint,12pt]{elsarticle}

\usepackage{amssymb}
\usepackage{graphicx}
\usepackage{rotating}
\usepackage{mathrsfs}
\usepackage{amsmath}
\usepackage{lscape}
\journal{Nuclear Physics A}

\begin{document}

\begin{frontmatter}



\title{Analysis of the unbound spectrum of $^{12}$Li}


\author[label1]{Z.~X.~Xu}

\author[label1]{R.~J.~Liotta}

\author[label1]{C.~Qi\corref{email}}

\author[label2]{T.~Roger}

\author[label2]{P.~Roussel-Chomaz}

\author[label2]{H.~Savajols}

\author[label1]{R.~Wyss}

\address[label1]{Royal Institute of Technology (KTH), Alba Nova University Center,
SE-10691 Stockholm, Sweden}
\address[label2]{GANIL, CEA/DSM - CNRS/IN2P3, Bd Henri Becquerel, BP 55027, F-14076 Caen Cedex 5, France}

\cortext[email]{Corresponding author.\\
\textit{E-mail address:} chongq@kth.se (Chong Qi)}

\begin{abstract}
The unbound nucleus  $^{12}$Li is evaluated by studying three-neutron
one-proton excitations within the multistep shell
model in the complex energy plane.
It is found that the
ground state of this system consists of an antibound $2^-$ state. A number
of narrow states at low energy are found which ensue from  the coupling 
of resonances in $^{11}$Li to continuum states close to 
threshold. 

\end{abstract}


\end{frontmatter}

% \linenumbers




%% References

\section{Introduction}

The study of halo nuclei is one of the main subjects of research in nuclear
physics at present. Many theoretical predictions on halo, superhalo and
antihalo nuclei have been advanced in recent years~\cite{hal1,hal2,hal3}.
Most of these calculations
correspond to nuclei very far from the stability line. They are
mainly thought as a guide for experiments to be performed in coming facilities.
The general feature found in these calculations is that a necessary
condition for a nucleus to develop a halo is that the outmost nucleons move
in shells which extend far in space. That is, only a weak barrier keep the
system within the nuclear volume. These shells may be resonances, antibound
states (also called virtual states), or even low-spin bound states which lie very
close to the continuum threshold.
These conditions are fulfilled by the nucleus $^{11}$Li and also heavier Li
isotopes. There are a number of experiments which have been performed in
these very unstable isotopes in order to get information about the structure
of halos \cite{bau07}. In particular we will concentrate our attention to Refs.
\cite{aks08,pat10,hal10} where the spectrum of $^{12}$Li was measured. Our aim is
to analyze these experimental data by using a suitable formalism to treat
unstable nuclei. This formalism is an extension of the shell model to the
complex energy plane and is therefore called complex shell model
\cite{idb02}, although the name Gamow shell model is also used \cite{mic02}.
In addition, the correlations induced by the pairing force acting upon
particles moving in decaying single-particle states will be taken into
account by using the multistep shell model (MSM) \cite{lio82}.

The formalism is presented in Section \ref{form}. Applications are in
Section \ref{appl} and a summary and conclusions are in Section \ref{sumc}.

\section{The formalism}
\label{form}
The study of unstable nuclei is a very difficult undertaking since, in
principle, time dependent formalisms should be used to describe the motion
of a decaying nucleus. However, the system may be considered stationary
if it lives a long time. In this case the time dependence can be circumvented.
In fact, often unstable nuclei live a very long time and therefore they may be
considered bound as, e.g., in alpha decaying states of many heavy isotopes, like
$^{208}$Bi or $^{180}$Ta(9$^-$), with $T_{1/2}>10^{15} y$.
On the other hand, experimental
facilities allow one nowadays to measure systems living a very short time. To
describe these short time processes one has
to consider the decaying character of the system.

Of the various theories
that have been conceived to analyze unbound systems, we will apply an extension
of the
shell model to the complex energy plane~\cite{idb02}. The basic assumption
of this theory is that resonances can be described in terms of states
lying in the complex energy plane. The real parts of the corresponding
energies are the positions of the resonances
while the imaginary parts are minus twice the corresponding widths, as it was
proposed by Gamow at the beginning of quantum mechanics~\cite{gam28}.
These complex states correspond to solutions of the Schr\"odinger equation with
outgoing boundary conditions. We will not present here the formalism in
detail, since this was done many times before, e.g., in Refs.
\cite{cxsm,mic09}. Rather, we will give the main points necessary for the
presentation of the applications.

\subsection{The Berggren representation}
\label{sec:berg}

In this Subsection we will very briefly describe the representation to be
used here.

The eigenstates of a central potential obtained as outgoing
solutions of the Schr\"odinger equation can be used to
express the Dirac $\delta$-function as \cite{b68},
\begin{equation}\label{eq:delb}
\delta(r-r')=\sum_n w_n(r) w_n(r') + \int_{L^+} dE u(r,E) u(r',E),
\end{equation}
where the sum runs over all the bound and antibound states plus the complex
states
(resonances) which lie between the real energy axis and the integration contour
$L^+$.  The wave function of a state $n$ in these discrete set is
$w_n(r)$ and  $u(r,E)$ is the scattering function at energy $E$.
The antibound states are virtual states with negative scattering length.
They are fundamental to describe nuclei in the Li region \cite{thz}.

The resonances and the antibound states are poles of the single-particle
Green function and, therefore, we will call them "poles" in order to make a
distinction with the scattering states. This is important since it is
through the poles that we will recognize physically meaningful states, i. e.
states that live a time long enough. Although meaningful states are 
usually immersed among continuum states, they can be recognized because
their wave functions contain important contributions from the poles, as will
be seen in the Applications.

Discretizing the integral of Eq. (\ref{eq:delb}) one obtains
the set of orthonormal vectors $\vert \varphi_j\rangle$
forming the Berggren representation \cite{lio96}. Since this discretization
provides an approximate value of the integral, the Berggren vectors fulfill
the relation
$I\approx\sum_j \vert \varphi_j\rangle \langle \varphi_j\vert$,
where all states, that is bound, antibound, resonances  and discretized
scattering states, are included. The corresponding single-particle wave
functions are
\begin{equation}\label{eq:spwf}
\langle \vec r\vert \varphi_i\rangle
=R_{n_il_ij_i}(r)\big (\chi_{1/2}Y_{l_i}({\hat r})\big )_{j_im_i},
\end{equation}
where $\chi$ is the spin wave function and
\begin{equation}\label{eq:rphi}
R_{n_il_ij_i}(r)
=\phi_{n_il_ij_i}(r)/r
\end{equation}
is the radial wave function fulfilling the Berggren metric, according to which the
scalar product between two functions consists of one function times the other
(for details see Ref. \cite{lio96}), i.e.,
\begin{equation}\label{eq:bergm}
\int_0^{\infty} dr \phi_{n_il_ij_i}(r)\phi_{n'_i l_ij_i}(r)
=\delta_{n_in_i'}.
\end{equation}

We will apply the Berggen representation to analyze the spectrum of
$^{12}$Li by using the Multistep Shell Model Method \cite{lio82} (MSM).
In order to make the presentation clear we will give a short description
of the main points to be used in this paper.

The Berggren representation has been used before within the framework of the 
CXSM (or the Gamow Shell Model, which is the same \cite{mic09}) to study Li 
isotopes \cite{mic04}.
The core in these calculations was assumed to be $^{4}$He and the Shell
Model single-particle states were the shells $0p$ only.
As  pointed out in that reference, the study of
many particles moving in states lying in the complex energy plane can be a
challenging task. The main problem is that due to the presence of the
scattering states the dimension of the Berggren basis soon becomes very
large, as well as non-Hermitian and complex.
Therefore in \cite{mic04} one could describe well light Li isotopes,
up to $^9$Li. But in heavier isotopes  the state $1s_{1/2}$ (which is an
antibound state) is of a fundamental importance and its inclusion would make
the application of the CXSM a
prohibitive undertaking \cite{mic04}. The importance of the antibound state
in this nuclear region was known since a rather
long time \cite{thz}. An excellent explanation of the halo nucleus $^{11}$Li
was given in Ref. \cite{ber91} by using the Continuum Shell Model, including
in the representation only the neutron waves
$s_{1/2}$, $p_{1/2}$ and $d_{3/2}$, which in
terms of the CXSM implies to include the antiboud state and the resonances
$0p_{1/2}$ and $0d_{3/2}$, but not the state $0p_{3/2}$. The corresponding CXSM
calculation was indeed performed in Refs. \cite{idb04,ant} and, perhaps not
surprising, one could thus explain well the structure of $^{11}$Li as well as the
corresponding halo.

One way of
avoiding too large dimensions is by including in the basis only physically
meaningful states. These are states which govern the calculated quantities.
For instance, in the
evaluation of $^6$Li the single-particle resonance $0p_{1/2}$ is very broad
and its inclusion does not affect the results appreciably \cite{mic04}.
In this context, one important point that has to be
emphasized is the evolution of the single-particle states in these isotopes
as the continuum threshold is reached. In light Li isotopes the unstable
character of the unbound states is reflected by the CXSM in that the
neutron resonance $0p_{1/2}$ is very broad and plays practically no role in
the evaluation of the spectrum. Instead, starting in $^9$Li this state
becomes extremely important and in $^{11}$Li contributes by about 50\% to the
formation of the halo \cite{ant,gar}. On the other hand, the state $0p_{3/2}$
is fundamental in the structure of light Li isotopes, but its importance is
diminished to a point where it can be neglected starting in $^{10}$Li. Shell
Model calculations performed within this approximation reproduced very well
the experimental data \cite{ber91,ant,esb97}. Even three-body (Fadeev) approaches
\cite{gar,Shu09} have shown that the structure of heavy Li isotopes are not
appreciably affected by the state $0p_{3/2}$.

An important conclusion of all the calculations mentioned above
is that the low lying
states in $^{11}$Li are essentially two-neutron excitations. Moreover,
the corresponding single-neutron states, forming the CXSM representation,
are in a similar way extracted from the odd-neutron excitations in
$^{10}$Li \cite{ant,gar}. It is to be noticed that in these references the proton
degrees of freedom were also neglected. The reason for this is that
the pairing interaction acting upon the even number of
neutrons present in odd Li isotopes determines the spectrum, leaving the
protons as mere spectators. However, this cannot be said in our case of
$^{12}$Li. We will therefore include in the calculations the proton
shell $0p_{3/2}$.

In our three-neutron one-proton case a way of distinguishing physically
meaningful states is by considering first the two-neutron  excitations.
From
the calculated states one can single-out the resonances which are very broad
or otherwise unphysical (for instance energy eigenvalues with real as well as
imaginary negative parts).

In order to introduce the physically meaningful
two-neutron states in the formalism
we will apply the MSM. In this method one
solves the Shell Model equations in several steps. One first chooses a
single-particle representation. Then one evaluates the
two-body equations. Next one solves the three-body equations within a basis
consisting of the tensorial product of the one- and two-body basis
previously evaluated. For the four-particle case one can choose as a basis
the one- times three-particle basis states already evaluated or the two- times
two-particle basis, and so on. In our case we will first solve the
two-neutron states as done in Refs. \cite{ant}. We will thus sort out the
physical meaningful states to be used in the three-neutron case. Besides the
advantage of reducing the basis dimensions, the inclusion of these
meaningful two-particle states implies that relevant continuum
(scattering) states, that is those continuum states that determine the
physically meaningful two-particle resonances, are also included.

In the next MSM step we will evaluate the three-neutron states in a basis
consisting of the one- times two-particle states.
Finally, with the three-neutron states thus determined we will form the
one-proton three-neutron MSM basis to evaluate the spectrum of $^{12}$Li.

The formalism corresponding to these calculations starts by choosing
the single-particle (Berggren) states.  Using the Berggren representation
thus chosen one gets the
two-particle shell-model equations in the complex energy plane (CXSM)
\cite{cxsm}, i.e.,
\begin{equation}\label{eq:sme}
(W(\alpha_2)-\epsilon_i-\epsilon_j)X(ij;\alpha_2)=
\sum_{k\leq l}\langle\tilde k\tilde l;\alpha_2\vert V\vert ij;\alpha_2\rangle X(kl;\alpha_2),
\end{equation}
where $V$ is the residual interaction. The tilde in the interaction matrix
element denotes mirror states so that in
the corresponding radial integral there is not any complex conjugate,
as required by the Berggren metric.
The two-particle states are labeled by
$\alpha_2$ and Latin letters label single-particle
states.
$W(\alpha_2)$ is the correlated two-particle energy and $\epsilon_i$
is  single-particle energy. The two-particle
wave function is given by
\begin{equation}\label{eq:wfsq}
\vert \alpha_2\rangle=P^+(\alpha_2)\vert 0\rangle,
\end{equation}
where the two-particle creation operator is given by,
\begin{equation}\label{eq:wfsq}
P^+(\alpha_2)=\sum_{i\leq j}X(ij;\alpha_2)
\frac{(c^+_ic^+_j)_{\lambda_{\alpha_2}}}{\sqrt{1+\delta_{ij}}},
\end{equation}
and $\lambda_{\alpha_2}$ is the angular momentum
of the two-particle state.

We will use a separable interaction, as in Ref. \cite{ant}, which
describes well the two-neutron states in $^{11}$Li.
The energies are thus obtained by solving the corresponding
dispersion relation.
The two-particle wave function amplitudes are given by \cite{ant}
\begin{equation}
\label{eq:tpwf}
 X(ij;\alpha_2) = N_{\alpha_2} \frac{f(ij,\alpha_2)}{\omega_{\alpha_2}-
(\epsilon_i + \epsilon_j)},
\end{equation}
where $f(ij,\alpha_2)$ is the single particle matrix element of the field
defining the separable interaction and
$N_{\alpha_2}$ is the normalization constant determined by the condition
$\sum_{i \le j} X(ij;{\alpha_2})^2 = 1$.

With the two-neutron states thus evaluated we proceed to the calculation of
the three-neutron states by using the MSM in the complex energy plane
(CXMSM), as briefly described below.

\subsection{The Multistep Shell Model Method}
\label{sec:msm}
As its name indicates, the Multistep Shell Model Method (MSM) solves the
shell model equations in several steps. In the first step the
single-particle representation is chosen. In the second step the energies
and wave functions of the two-particle system are evaluated by using a given
two-particle interaction. The three-particle states are evaluated in terms
of a basis consisting of the tensorial product of the one- and two-particle
states previously obtained. In this step the interaction
does not appear explicitly in the formalism. Instead, it is the wave functions
and energies of the components of the MSM basis that replace the
interaction. The MSM basis is overcomplete and non-orthogonal.
To correct this one needs to evaluate the overlap matrix among the basis
states.
A general description of the formalism is in Ref. \cite{lio82}.
The particular system that is of our interest here, i.e., the
three-particle case, can be found in Ref. \cite{blo84}, where the MSM was applied
to study the three-neutron hole states in the nucleus $^{205}$Pb.

Using the Berggren single-particle representation described above, we will
evaluate the complex
energies and wave functions of $^{12}$Li using the MSM basis states
consisting of the Berggren one-particular states, which are states in
$^{10}$Li, times the two-neutron excitations that determine the low lying
spectrum of  $^{11}$Li. Below we
refer to this formalism as CXMSM.

The three-particle energies $W(\alpha_{3})$ are given by \cite{blo84}
\begin{multline}\label{TDA3}
    (W(\alpha_{3})-\varepsilon_{i}-W(\alpha_{2}))\langle\alpha_{3}|(c^{+}_{i}P^{+}(\alpha_{2}))_{\alpha_{3}}|0\rangle\\
    =\sum_{j\beta_{2}}\left\{\sum_{k}(W(\beta_{2})-\varepsilon_{i}-\varepsilon_{k})A(i\alpha_{2},j\beta_{2};k)\right\}
    \langle\alpha_{3}|(c^{+}_{j}P^{+}(\beta_{2}))_{\alpha_{3}}|0\rangle,
\end{multline}
where Latin letters label single-particle states and $\varepsilon$ are the
corresponding single-particle energies. The n-particle correlated states are
labelled by Greek letters with the subindex $n$. For instance $\alpha_3$
($\beta_2$), is a three (two) particle correlated state carrying energy
$W(\alpha_3)$ ($W(\beta_2)$). The function $A$ is
\begin{equation}
    A(i\alpha_{2},j\beta_{2};k)=\hat{\alpha}_{2}\hat{\beta}_{2}Y(kj;\alpha_{2})Y(ki;\beta_{2})\left\{\begin{array}{ccc}
                                                                        i & k & \beta_{2} \\
                                                                        j & \alpha_{3} & \alpha_{2}
                                                                      \end{array}
    \right\},\\
\end{equation}
and
\begin{equation}
    Y(ij;\alpha_{2})=(1+\delta(i,j))^{1/2}X(ij;\alpha_{2}).
\end{equation}

The matrix defined in Eq. (\ref{TDA3}) is not hermitian and the dimension
may be larger than the corresponding
shell-model dimension. This is due to the violations of the Pauli principle as well as overcounting of
states in the CXMSM basis.
Therefore the direct diagonalization of Eq. (\ref{TDA3}) is not convenient.
One needs to calculate the overlap matrix in order to transform the CXMSM basis into an orthonormal set.
In this three-particle case the overlap matrix is
\begin{equation}\label{Overlap}
    \left\langle0|(c^{+}_{i}P^{+}(\alpha_{2}))^{\dag}_{\alpha_{3}}(c^{+}_{j}P^{+}(\beta_{2}))_{\alpha_{3}}|0\right\rangle
    =\delta_{ij}\delta_{\alpha_{2}\beta_{2}}+\sum_{k}A(i\alpha_{2},j\beta_{2};k).
\end{equation}

Using the overlap one can transform the matrix determined by Eq. (\ref{TDA3})
into a hermitian matrix $T$ which has the right dimension.
The diagonalization of $T$ provides the three-particle energies. The
corresponding wave function
amplitudes can be readily evaluated to obtain
\begin{eqnarray}\label{alpha3}
    |\alpha_{3}\rangle&=&P^{+}(\alpha_{3})|0\rangle,\\
    P^{+}(\alpha_{3})&=&\sum_{i\alpha_{2}}X(i\alpha_{2};\alpha_{3})(c^{+}_{i}P^{+}(\alpha_{2}))_{\alpha_{3}},
\end{eqnarray}
where $P^{+}(\alpha_{3})$ is the three-particle creation operator.

It has to be pointed out that in cases where the basis is overcomplete the amplitudes
$X$ are not uniquely defined and, therefore, they do not have physical meaning,
Instead, the projection of the basis vector upon the corresponding
physical vector, i.e.,
\begin{equation}\label{funf}
F(i\alpha_2;\alpha_3) =  \left\langle\alpha_{3}|(c^{+}_{i}P^{+}(\alpha_{2}))_{\alpha_3}|0\right\rangle,
\end{equation}
is well defined.

That the CXMSM wave function amplitudes are
usually not well defined  is no hinder to evaluate the physical quantities.
For details see Ref. \cite{blo84}. It has to be pointed out that this
feature does not appear  when  the basis is
orthonormal, as in the one-proton three-neutron case to be analyzed below,
since neither the Pauli principle nor overcounting of states are relevant
here.

The next step in our CXMSM is the evaluation of the three-neutron one-proton
states. With the basis denoted as
\begin{equation}
|p\alpha_3;\alpha_4\rangle=(c^+_p P^+(\alpha_3))_{\alpha_4}|0\rangle,
\end{equation}
where $p$ labels the proton state, $\alpha_3$ is as in Eq. (\ref{alpha3}) and
$\alpha_4$ are the three-neutron
one-proton state,  the four-particle energies $W(\alpha_4$) in $^{12}$Li are
given by
\begin{multline}\label{TDA4}
(W(\alpha_4)-\varepsilon_p-W(\alpha_3))\langle\alpha_4|(c^+_pP^+(\alpha_3))_{\alpha_4}|0\rangle\\
\begin{split}
    =&\sum_{q\beta_3}\left\{\sum_{kl\lambda\alpha_2}\langle pk;\lambda|V|ql;\lambda\rangle B_1+\sum_{ijkl\lambda\alpha_2\beta_2}\langle pi;\lambda|V|ql;\lambda\rangle B_2\right\}\\
    &\times\langle\alpha_4|(c^+_qP^+(\beta_3))_{\alpha_4}|0\rangle,
\end{split}
\end{multline}
where,
\begin{multline}
    B_1=(-1)^{p+q+k+l}X(k\alpha_2;\alpha_3)F(l\alpha_2;\beta_3)\\\times\hat{\alpha}_3\hat{\beta}_3\hat{\lambda}^2\left\{
                                                                  \begin{array}{ccc}
                                                                    p & k & \lambda \\
                                                                    \alpha_2 & \alpha_4 & \alpha_3 \\
                                                                  \end{array}
                                                                \right\}\left\{
                                                                  \begin{array}{ccc}
                                                                    q & l & \lambda \\
                                                                    \alpha_2 & \alpha_4 & \beta_3 \\
                                                                  \end{array}
                                                                \right\},
\end{multline}
and
\begin{multline}
    B_2=(-1)^{p+q+i+l}Y(ji;\alpha_2)Y(jk;\beta_2)X(k\alpha_2;\alpha_3)F(l\beta_2;\beta_3)\\
        \times\hat{\alpha}_2\hat{\alpha}_3\hat{\beta}_2\hat{\beta}_3\hat{\lambda}^2\left\{
                                                                                             \begin{array}{ccc}
                                                                                               p & i & \lambda \\
                                                                                               \beta_2 & \alpha_4 & \alpha_3 \\
                                                                                             \end{array}
                                                                                           \right\}\left\{
                                                                                             \begin{array}{ccc}
                                                                                               q & l & \lambda \\
                                                                                               \beta_2 & \alpha_4 & \beta_3 \\
                                                                                             \end{array}
                                                                                           \right\}\left\{
                                                                                             \begin{array}{ccc}
                                                                                               i & j & \alpha_2 \\
                                                                                               k & \alpha_3 & \beta_2 \\
                                                                                             \end{array}
                                                                                           \right\}.
\end{multline}
Here $p$ and $q$ label proton states, while $i,j,k,l$ label neutron states.
The proton-neutron interaction matrix elements
$\langle pk;\lambda|V|ql;\lambda\rangle$ will be discussed in the
Applications. The wave function amplitudes $X$ and the projected quantities
$F$, defined above (Eq. (\ref{funf})), have been evaluated in previous steps
of the CXMSM.
Notice that in this case the overlap matrix is the unit matrix, i.e.,
\begin{equation}
\langle 0|(c^+_{p'} P^+(\alpha'_3))^\dag_{\alpha_4}(c^+_p P^+(\alpha_3))_{\alpha_4}
|0\rangle =\delta_{pp'}\delta_{\alpha_3\alpha'_3}
\end{equation}

The advantage of the MSM in stable nuclei
is that one can study the influence of collective
vibrations upon nuclear spectra within the framework of the shell model.
Thus, in Ref. \cite{blo84} the multiple structure of particle-vibration
coupled states in  odd Pb isotopes was analyzed.
But the most important
feature for our purpose is that the CXMSM allows one
to choose in the basis states
a limited number of excitations. This is because in the continuum
the vast majority of basis states consists of scattering functions. These do
not affect greatly physically meaningful two-particle states. That is, the
majority of the two-particle states provided by the CXSM are complex
states which form a part of the continuum background. Only a few of those
calculated states are relevant, namely the ones that are mainly built
upon poles.
The question of how to evaluate and recognize the
physically meaningful three-particle states, are addressed in the next
Section.

\section{Applications}
\label{appl}

In this Section we will apply the CXMSM formalism described above to study
the nucleus $^{12}$Li.

To evaluate the valence shells we will proceed as in Refs.
\cite{idb04,esb97,pac02} and choose as central field a Woods-Saxon potential
with different depths for even and odd orbital angular momenta $l$.
The corresponding parameters are (in parenthesis for odd $l$-values)
$a$= 0.670 fm, $r_0$ = 1.27 fm, $V_0$ = 50.0 (36.9) MeV and $V_{\rm so}$=16.5 (12.6)
MeV. As in Ref. \cite{idb04}, we thus found the single-particle bound states
$0s_{1/2}$ at -23.280 MeV  and $0p_{3/2}$ at -2.589 MeV.
The valence shells are the low lying resonances $0p_{1/2}$ at
(0.195,-0.047) MeV and $0d_{5/2}$ at (2.731, -0.545) MeV and the shell
$0d_{3/2}$ at (6.458,-5.003) MeV. This cannot be considered a resonance,
since it is so wide that rather it is a part of the continuum background.
Besides, the state $1s_{1/2}$ appears as an
antibound state at -0.050 MeV. We thus reproduce the experimental
single-particle energies as given in Ref. \cite{boh97}.
We also found other states  at higher energies, but they do not affect our
calculation because they are very high and also very wide.
We thus include in our Berggren representation only the antibound state
$1s_{1/2}$ and the resonances $0p_{1/2}$ and $0d_{5/2}$.

To define the Berggren single-particle representation we still have to
choose the integration contour ${L^+}$ (see Eq. (\ref{eq:delb})).

To include in the representation the antibound $1s_{1/2}$ state as well as the
Gamow resonances $0p_{1/2}$ and $0d_{5/2}$ we will use two different contours.
The number of points on each contour define the energies of the scattering
functions in the Berggren representation, i.e., the number of basis states
corresponding to the continuum background. This number is not uniformity
distributed, since in segments of the contour which are close to
the antibound state or to a resonance the scattering functions increase
strongly. We therefore chose the density of points to be larger in those
segments.

\begin{figure}[htdp]
\centerline{\includegraphics[width=0.6\textwidth]{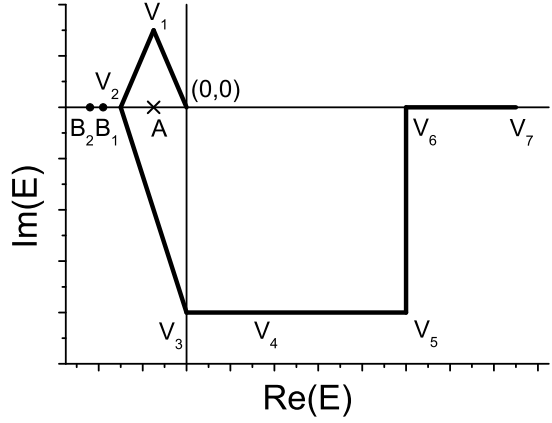}}
\caption{Contour used to include the antibound state (see, also, Ref.~\cite{idb04}). The points
$B_i$ denote bound states  while
$A$ denotes the antiboud state.
The points $V_i$ correspond to the vertices defining the contour.
They have the values $V_1$=(-0.05,0.05) MeV, $V_2$=(-0.1,0) MeV,
$V_3$=(0,-0.4) MeV, $V_4$=(0.5,-0.4) MeV, $V_5$=(8,-0.4) MeV, $V_6$=(8,0) MeV and
$V_7$=(10,0) MeV}
\label{cont}
\end{figure}

\begin{figure}[htdp]
\centerline{\includegraphics[width=0.6\textwidth]{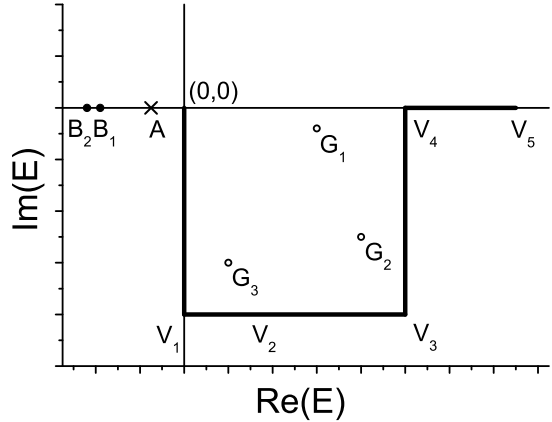}}
\caption{Contour used to include the Gamow resonances represented by the
points $G_i$. The vertices are $V_1$=(0,-1) MeV,
$V_2$=(1,-1) MeV, $V_3$=(8,-1) MeV, $V_4$=(8,0) MeV and $V_5$=(10,0) MeV.}
\label{contr}
\end{figure}

\begin{landscape}
\begin{table}
  \centering
  \caption{Number of Gaussian points in the different segments of the
contour of Fig. \ref{cont}.}\label{ngp}
  \begin{tabular}{cccccccc}
    \hline
Segment & [$(0,0)-V_1$]& [$V_1-V_2$]& [$V_2-V_3$]&[$V_3-V_4$]&[$V_4-V_5$]& [$V_5-V_6$]&[$V_6-V_7$]\\
Number  & 30           & 30         & 30         & 30        & 30        &    16        &   6     \\
    \hline
  \end{tabular}
\end{table}

\begin{table}
  \centering
  \caption{Number of Gaussian points in the different segments of the
contour of Fig. \ref{contr}.}\label{ngr}
  \begin{tabular}{cccccc}
    \hline
Segment & [$(0,0)-V_1$]& [$V_1-V_2$]& [$V_2-V_3$]&[$V_3-V_4$]& [$V_4-V_5$]\\
Number  & 30           & 30         & 30         & 8        &    4      \\
    \hline
  \end{tabular}
\end{table}
\end{landscape}

We include the antibound state by using the contour in
Fig. \ref{cont}.
The number of points in each segment are given in Table \ref{ngp}.
For the Gamow resonances the contour in Fig.
\ref{contr} is used with the number of Gaussian points as in Table
\ref{ngr}.

We have adopted these points after verifying that the results converged to
their final values.
A discussion about the choice of these contours and also on the physical
meaning of the antibound state can be found in Ref.~\cite{ant}.

\begin{figure}[htdp]
\centerline{\includegraphics[width=0.6\textwidth]{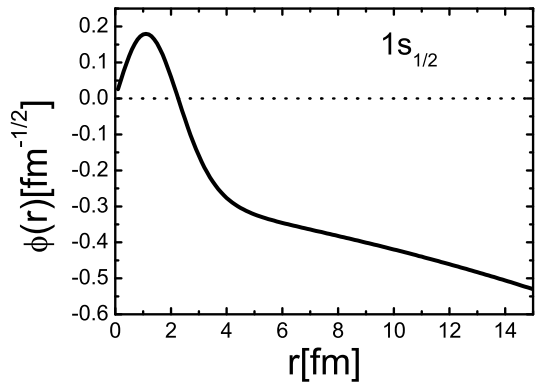}}
\caption{Radial function $\phi(r)$ corresponding
to the single-particle neutron antibound state $0s_{1/2}$ at an energy
of -0.050 MeV.}
\label{spwfs}
\end{figure}

\begin{figure}[htdp]
\centerline{\includegraphics[width=0.6\textwidth]{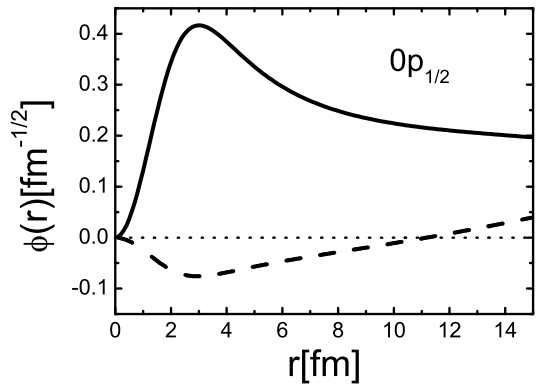}}
\caption{As Fig. \ref{spwfs} for the Gamow resonance
$0p_{1/2}$ at an energy of (0.195,-0.047) MeV. The dashed line
is the imaginary part of the wave function.}
\label{grp}
\end{figure}

\begin{figure}[htdp]
\centerline{\includegraphics[width=0.6\textwidth]{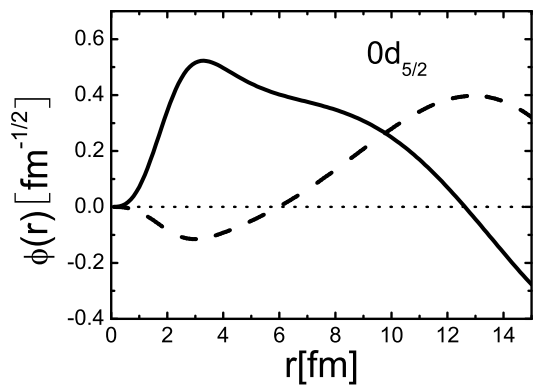}}
\caption{As Fig. \ref{spwfs} for the Gamow resonance
$0d_{5/2}$ at an energy of (2.731,-0.545) MeV. The dashed line
is the imaginary part of the wave function.}
\label{grd}
\end{figure}

To explore the extend to which the poles are physically meaningful we
plotted in Fig. \ref{spwfs} the $1s_{1/2}$ antibound state. One sees that it 
extends in an increasing rate far out from the
nuclear surface, as expected in this halo nucleus (the standard
value of the radius is here $1.2\times 11^{1/3}$=2.7 fm).  The radial wave
function corresponding to the Gamow resonance $0p_{1/2}$ is shown in Fig.
\ref{grp}.
The resonance $0d_{5/2}$ has a large and increasing imaginary part
at relative short distances, as shown in Fig. \ref{grd}.

With the single-particle representation thus defined we
proceed to evaluate the two-particle states.

\subsection{Two-particle states: the nucleus $^{11}$Li.}
We would like to start this Subsection by stressing, once again, that the
two-particle states that we will investigate here correspond to two-neutron
excitations in the spectrum of $^{11}$Li, while the corresponding odd proton
is an spectator. To avoid confusions in what follows we will call these states
$^{11}$Li($2\nu;J$), where $2\nu$ indicates that the state $J$, which belongs to
the spectum of $^{11}$Li, is determined by two-neutron excitations.

The only state which is measured in $^{11}$Li is its bound ground state, which
was  found to lie at an energy of -0.369 MeV \cite{smi08}. The
corresponding angular momentum is $3/2^-$. This spin arises from the odd
inert proton,
lying deep in the spectrum, coupled to two neutrons.
The dynamics of the system is thus determined by the pairing
force acting upon the two neutrons coupled to a state $0^+$, which behaves
as a normal even-even ground state \cite{ant,esb97}.
Besides the energy, this state has been measured to have an angular momentum
contain  of about 60\% of  s-waves and 40\% of p-waves,
although small components of
other angular momenta are not excluded \cite {gar}.

We will perform the calculation of the two-particle states by using the
separable interaction discussed in Section \ref{form}. The
strength $G_{\lambda_2}$, corresponding to the states with
angular momentum $\lambda_2$ and parity $(-1)^{\lambda_2}$, will be
determined by fitting the experimental energy of the lowest of these states,
as usual. It is worthwhile to point out that $G_{\lambda_2}$
defines the Hamiltonian and, therefore, is a real quantity. The two-particle
energies are found by solving the corresponding dispersion relation while the
two-particle wave function components are as in Eq. (\ref{eq:tpwf}).

With the quantities entering the two-particle TDA equations thus determined
we evaluated the two-neutron states in $^{11}$Li($2\nu;J=0^+$). We found that
the angular momentum contain of the ground state wave function is
46.8\% $s$-states, 49.1\% $p$-states and 4.2\% $d$-states. This is
in reasonable agreement with experiment \cite{gar}.

The wave function components corresponding to $^{11}$Li($2\nu;gs)$ are strongly
dependent upon the contour that one uses. However, measurable quantities,
like the energies and transition probabilities, do not. This is because the
physical quantities are defined on the real energy axis and, therefore, they
remain the same when changing contour. But complex states which are part of
the continuum background do not have any counterpart on the real energy axis
and the physical  quantities for these states acquire different values for
different contours~\cite{ant}. We will use this property to determine
whether a complex state is a meaningful resonance.  This is important, since
the ground state is the only one for which experimental data exists. There
might be other meaningful states that have not been found yet. This implies
that we have to evaluate all possible two-particle states which are spanned
by our single-particle representation. For the purpose of this paper this is
an important task, since in the next step of the CXMSM only physically
meaningful states will be considered as members of the basis.

To decide whether a calculated state is a meaningful resonance we will
proceed as in Refs. \cite{ant,gpr} and analyze the singlet (S=0) component
of the two-particle wave function.
The corresponding expression for this component  was
given in Eq. (10) of Ref. \cite{gpr}, but we will show it here again for
clarity of presentation. For the
state $\alpha$ with spin and spin-projection $(JM)$ that component is,
with standard notation,
\begin{multline}\label{eq:wfs0}
 \Psi_{\alpha JM}(\vec{r}_1 \vec{r}_2) =
    \left[ \chi_{1/2}(1) \chi_{1/2}(2) \right]_0^0
    \sum_{a \le b} X(ab,\alpha JM)\hat{j}_a \hat{j}_b\\ \times
    \left[ C(ab,\vec{r}_1 \vec{r}_2) - (-)^{j_a+j_b-J} C(ba,\vec{r}_1 \vec{r}_2) \right],
\end{multline}
where
\begin{equation}
 C(ab,\vec{r}_1 \vec{r}_2)= \phi_a(r_1) \phi_b(r_2) (-)^{l_b+1/2-j_a+J}
  \left\{
   \begin{array}{ccc}
     l_a & j_a & 1/2 \\
     j_b & l_b & J \\
   \end{array}
  \right\}
  \left[ Y_{l_a}(\hat{r}_1) Y_{l_b}(\hat{r}_2) \right]_{JM},
\end{equation}
and $\phi_a(r)$ is the radial wave function corresponding to the
single-particle state $a$ (Eq. (\ref{eq:rphi})).

If the two-particle state $(\alpha JM)$ is a meaningful resonance then the
wave function above should be localized within a region extending not too
far outside the nuclear surface, and its imaginary part should not be too large
\cite{ant}. To study these features it is not necessary to go to all six
dimensions corresponding to the coordinates $\vec r_1$ and $\vec r_2$.
In fact it is enough to consider the coordinate $r$ given by
$\vec r_1=\vec r_2=\vec r=(0,0,r)$ which corresponds to the two particles
located at the same point and in the z-direction. For details see
\cite{gpr}. We will call this one-dimensional function
$\Psi_{\alpha JM}(r)$.

The evaluation of the $0^+$ states is a
relatively easy task, since in this case we have determined the strength
$G_{0^+}$ by fitting the experimental energy of $^{11}$Li($2\nu;gs)$. With this
value of the strength we calculated all the $0^+$ states and found that 
the vast majority of them are continuum states which belong to the
background. The relevant states are those which are built
mainly from the poles \cite{idb02,mic02}. Although these states may not be
in themselves physically meaningful resonances, they can influence
significantly the spectrum of $^{12}$Li. 

\begin{figure}[htdp]
\centerline{\includegraphics[width=0.6\textwidth]{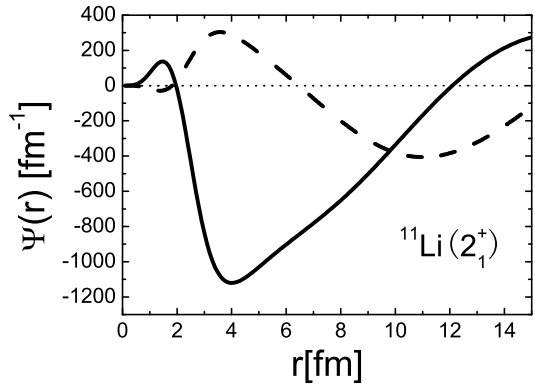}}
\caption{Radial function $\Psi(r)$ corresponding
to the two-particle state $^{11}$Li($2\nu; 2^+_1$) at an energy
of (2.300,-0.372)MeV. The dashed line
is the imaginary part of the wave function.}
\label{wf2p}
\end{figure}

We therefore included in the CXMSM basis those two-particle states which 
have at least one pole configuration which, in
absolute value, is 0.3 or larger. We found that all these states correspond
to the poles $1s_{1/2}$ and $0p_{1/2}$ coupled to themselves or to continuum
states lying close to threshold. As a result, these are narrow states.
We found also similar meaningful $1^-$ and $2^+$ states.
Specially important is the state $1^-$ at  (0.084,-0.002) MeV which is
built by the pole configuration $(0s_{1/2}0p_{1/2})_{1^-}$ and therefore
may be physically meaningful.  Also the  
state $2^+_1$ at (2.300,-0.372) MeV which is practically built by the pole
configuration $(0s_{1/2}0d_{5/2})_{2^+}$ can be meaningful. However its width 
(i. e. 0.744 MeV) seems to be too large and therefore we decided to analyze 
the corresponding radial wave function in more detail. For this we drew
$\Psi_{2^+_1}(r)$, as shown in
Fig. \ref{wf2p}.   One sees that the wave function
is rather localized and that its imaginary
part is relatively small as
compared to the corresponding real part. This is a state which perhaps is at
the limit of what can be considered a meaningful resonance.
Yet, it has an effect on the physical
three-particle states, as will be seen below. It is worthwhile to point out
that the width of this state (744 keV) is the escape width. At high
energies, where the giant resonances lie, most of the width consists of the
spreading width, i.e., of mixing with particle-hole configurations
\cite{gal88}. However, at the low energies of the states that we study this
mixing is not relevant.

An important point for the analysis of the three-particle states to be
performed below, is that the scattering wave functions in the segments
$[(0,0)-V_1]$, $[V_1-V_2]$ and $[V_2-V_3]$ are similar in magnitude to the
wave function of the antibound state. This is because the segments are very
close to the antibound state \cite{ant}. This is a feature that cannot be
avoided, and is due to the attractive character of the pairing force. That
is, the lowest single-particle configuration in the Berggren basis is
$V_2^2$, with energy $-2\epsilon$, where $V_2=(-\epsilon,0)$. This
configuration has to lie {\it above} the energy of the two-particle
correlated state, i.e., it has to be $\epsilon>\omega(^{11}$Li($2\nu;gs$))/2.

With the states $0^+$, $1^-$ and $2^+$ thus calculated we proceeded to the
calculation of the three-particle system within the CXMSM.

\subsection{Three-particle states}

Using the single- and two-particle neutron states
discussed above, we formed all the possible
three-particle basis states.
Due to the large number of scattering states
included in the single-particle representation the dimension of the
three-particle basis is also large. The scattering states are needed in
order to describe these unstable states.

With the CXMSM basis thus
constructed we evaluated the dynamical matrix Eq. (\ref{TDA3}) and the overlap
Eq. (\ref{Overlap}). With these we formed the symmetric Hamiltonian matrix
which we diagonalized to obtain the three-neutron states.
We found that the lowest state is $1/2^+$ at (-0.381,+0.023) MeV. That is, 
the energy is
real and  negative. It is an antibound state, as it
is the $1s_{1/2}$ state itself. A manifestation of this is that the radial
wave function diverges at large distances.

With the three-neutron states thus calculated 
and the proton state $0p_{3/2}$ we
constructed the final CXMSM basis to describe the nucleus $^{12}$Li.

\subsection{Four-particle states}

In this case the one-proton three-neutron CXMSM basis is orthonormal and the
matrix (\ref{TDA4}) is already the Hamiltonian matrix. All the three-neutron
quantities appearing in Eq. (\ref{TDA4}) have been solved in the previous
step of the CXMSM. Instead, the single-proton energy and the proton-neutron
interaction matrix elements are quantities that we have still to determine.
The single-proton energy was not considered so far. When we evaluated the
ground state of $^{11}$Li($2\nu)$ we assumed that its energy was only determined by
the two-neutron excitations. The value of the energy was obtained by fitting
the strength of the neutron-neutron separable interaction, ignoring any effect
that the protons may have had.  In the analysis of
the spectrum of $^{11}$Li, including the wave functions, this is irrelevant,
since the assumption in those calculations was that the protons were only
spectators. In other words, the effect of the odd proton
was only an scaling of the energies.  A similar feature occurs in the
evaluation of the three-neutron one-proton states. As seen in Eq. (\ref{TDA4})
the value of the proton energy $\varepsilon_p$, where $p$ is the
proton state $0p_{3/2}$, only shift the spectrum, but the relative
energy between two given states is not affected. Again here the effect of
the proton degree of freedom regarding the single-particle energy
is to scale the whole spectrum of $^{12}$Li.

The determination of $\varepsilon_p$ is a difficult task
due to the energy renormalizations that our procedure implies. However, it
should be a real number, since the proton is a bound state. We
will, therefore, not intend to evaluate the absolute energies and only
discuss the $^{12}$Li energies relative to the corresponding ground state
energy. This is equivalent to take $\varepsilon_p$
as a real parameter that adjust the real part of the
ground state energy of $^{12}$Li.
The proton-neutron interaction matrix elements were taken from empirical effective interactions which are determined by fitting experimental data \cite{war92,bro01}.

With the quantities entering the Hamiltonian matrix (\ref{TDA4}) determined
as discussed above, we calculated the energies $W(\alpha_4)$ and the wave
function amplitudes
$\langle\alpha_4|(c^+_pP^+(\alpha_3))_{\alpha_4}|0\rangle$. Notice that in
this case of an orthonormal basis the wave function components $X$ and the
corresponding projections $F$ (Eq. (\ref{funf})) coincide.

Due to the presence of the continuum states  the
dimension of the one-proton three-neutron basis
is very large. Of all the states calculated within this basis we will present 
only those which are physically meaningful. As we have discussed above, the
main configurations in such states contain large contributions from the
poles (relevant configurations).  We therefore will proceed as in the
evaluation of the two-neutron states and consider physically 
meaningful the one-proton three-neutron basis states 
for which at least the amplitude of
one relevant configuration is, in absolute value, 0.3 or larger. 

Below we will briefly describe the structure of
the states thus calculated forming the $^{12}$Li spectrum.

\begin{figure}[htdp]
\centerline{\includegraphics[width=0.6\textwidth]{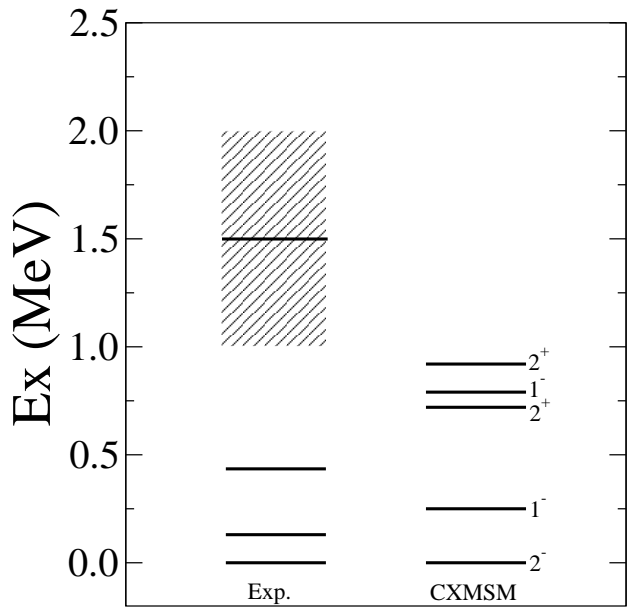}}
\caption{ Experimental level scheme in $^{12}$Li. The three lowest levels are
from \cite{hal10}, while the one at 1.5 MeV is from \cite{pat10}. The
theoretical results are labelled  CXMSM.}
\label{smen}
\end{figure}

As seen in  Fig. \ref{smen} there are five states at low energy predicted by 
the theory. All of them are narrow and mainly built upon configurations that 
are close
to the continuum threshold. This is not surprising since the determining
poles at low energy are the antibound state $s_{1/2}$ and the 
resonance  $p_{1/2}$, and both have real or nearly real energies and
are very close to threshold.. 

The $^{12}$Li ground state is $2^-$ and its energy is real. 
Since it is neither a bound state (its wave function
diverges at large distances) neither a resonance (its energy is a real
quantity) we conclude that it is an antibound state.
This is not surprising, since it is mainly built upon the CXMSM basis state 
$|[\pi (0p_{3/2})\nu (1s_{1/2}\otimes ^{11}{\rm Li(2\nu;gs)})
_{1/2^+}]{2^-}\rangle$,
where $\pi$ ($\nu$) indicates proton (neutron) degree of freedom.
The three-neutron component of this state is dominated by the
state $\nu (1s_{1/2})$ and this induces the state $2^-_1$ to be antibound.
It is worthwhile to point out that the 
influence of the $\nu (1s_{1/2})$ 
antibound single-particle state is due to its
position, lying very close
to the continuum threshold. Since there is no barrier to trap inside the
nucleus the
neutron moving in this state, the corresponding wave
function is very similar to the one corresponding to a bound state at the
same small and negative energy \cite{ant}. Within the Continuum Shell Model the
influence of the antibound state is taken into account by the scattering
states on the real energy axis
lying close to the continuum threshold \cite{ber98}. Even in the complex
energy plane the scattering wave functions lying close to the antibound
state are similar to each other \cite{ant,mic06}. A discussion on
the conveniences and drawbacks of using the complex energy plane to describe
the antibound state can be found in Ref. \cite{mic06}.

The question one may ask is how is it possible that in $^{12}$Li three 
neutrons can
occuppy the state $1s_{1/2}$. The answer to this is that  in
the continuum this state is split in many components.

The other states at low energy in Fig. \ref{smen} are built upon the proton
single-particle state $0p_{3/2}$ coupled to the three-neutron states 
$|1s_{1/2}\otimes ^{11}{\rm Li(2\nu;0^+)}\rangle$, 
$|1s_{1/2}\otimes ^{11}{\rm Li(2\nu;1^-)}\rangle$,
$|0p_{1/2}\otimes ^{11}{\rm Li(2\nu;0^)}\rangle$ and
$|0p_{1/2}\otimes ^{11}{\rm Li(2\nu;1^-)}\rangle$. They are all narrow
states.

Besides these states there are many levels that lie above 1 MeV which are
built upon the pole $0d_{5/2}$. This may explain the broad resonance seen 
experimentally at about 1.5 MeV.

The experiment shows two narrow excited states below 0.5 MeV, which 
the calculation predicts to be $1^-$ and $2^+$, respectively. 

Most of the calculated levels in Fig. \ref{smen} are strongly influenced by 
the continuum states. If we would take only relevant
configurations with amplitudes larger than 0.9 then only two states would
appear. If no continuum configurations are included then
no excited state below 1.1 MeV would be 
found, as can be seen in the shell model calculation of Ref. \cite{hal10}.

\section{Summary and conclusions}
\label{sumc}

In this paper we have studied excitations occurring in the continuum part of
the nuclear spectrum which are at the limit of what can be observed within
present experimental facilities. These states are very unstable but yet live
a time long enough to be amenable to be treated within stationary
formalisms. We have thus adopted the CXSM (shell model in the complex energy
plane \cite{cxsm}) for this purpose. In addition we performed the shell
model calculation by using the multistep shell model. In this method of
solving the shell model equations one proceeds in several steps. In each
step one constructs building blocks to be used in future steps \cite{lio82}.
We applied this formalism to analyze the spectrum of $^{12}$Li by assuming
that it is determined by one-proton three-neutron excitations.
First we studied the neutron degrees of freedom by profiting from the
information that exits on the single-neutron states in $^{10}$Li
and on two-neutron states in $^{11}$Li.
In our case of $^{12}$Li the neutron  excitations correspond to the motion of
three neutrons, partitioned as the one- times two-neutron systems.
This formalism was applied before, e.g., to study multiplets in the lead
region \cite{blo84}. Finally, the spectrum of $^{12}$Li is calculated by
coupling the three-neutron system with the $0p_{3/2}$ single-proton
state.

We adopted the single-particle energies (i.e., states in $^{10}$Li) as provided 
by experimental data when available or as provided by our calculation. These
are the antibound state $1s_{1/2}$ and the resonances $0p_{1/2}$ and 
$0d_{5/2}$. Besides these states (which are poles of the Green function) we
also included continuum states (Coulomb waves). We found
that the physically meaningful two-neutron states are mainly built upon
poles. With the two-particle states thus obtained we calcuated the
three-neutron states and coupled them to the $0p_{3/2}$ protons state to
evaluate the spectrum of $^{12}$Li,
The ground state energy of this nucleus turns out to be real.
Since this is neither a bound states (the wave function diverge at large
distances) nor a resonance (the energy us real) we conclude that it is an
antibound states. This agrees with a number of experiments 
\cite{aks08,pat10,hal10}.

Besides this antibound state our calculation predicts two low lying
narrow states which, as seen in Fig. \ref{smen}, are also found experimentally.
But there are also three other calculated levels which have not been
observed so far.

We  conclude that the experimentally observed narrow states in $^{12}$Li
arise as a result of the valence neutrons moving in  unstable shells.
As in $^{11}$Li, these are mainly the antibound state 
$1s_{1/2}$ and the narrow resonance $0p_{1/2}$. But in $^{12}$Li continuum
states lying close to threshold play also a fundamental role.

\section{Acknowledgments}

This work has been supported by the Swedish Research Council (VR). Z.X. is
supported in part by the China Scholarship Council under grant No.
2008601032.


%\bibliographystyle{elsarticle-num}






% \end{thebibliography}

\end{document}